\documentclass[a4]{paper}
\usepackage[utf8]{inputenc}
\usepackage[T2A]{fontenc}
\usepackage[english]{babel}
\usepackage{amssymb}
\usepackage{amsmath}
\usepackage{textcomp}
\usepackage{indentfirst}
\usepackage{float}
\usepackage{caption}
\usepackage{textcomp}
\DeclareCaptionLabelSeparator{pipe}{ $: $ }% or $\vert$
\usepackage{authblk}

\usepackage{units}
\usepackage{amsfonts}
\usepackage{graphicx} % ?????????? ????????
\usepackage{amstext} % ??? ??????????? ???????? ? ?????????????? ????????
\usepackage{wasysym} % ??? ??????????

\usepackage{comment}
\usepackage{color}

\usepackage{epstopdf}
\usepackage{xcolor}
\usepackage{hyperref}
 
\definecolor{linkcolor}{HTML}{000000} % ???? ??????
\definecolor{urlcolor}{HTML}{000000} % ???? ???????????

\numberwithin{equation}{section}

\hypersetup{pdfstartview=FitH,  citecolor=black, linkcolor=linkcolor,urlcolor=urlcolor, colorlinks=true}
\usepackage[left=2.5cm,right=2cm, top=2cm,bottom=2cm,bindingoffset=0cm]{geometry} % ????

\DeclareMathOperator{\sign}{sign}
\DeclareMathOperator{\Real}{Re}
\DeclareMathOperator{\Image}{Im}

\DeclareMathOperator{\Det}{Det}

\usepackage[sorting=none,backend=bibtex,citestyle=numeric]{biblatex}
\bibliography{literature.bib}
 
\begin{document}

\title{Linear Stability of Shock Waves in Ultrarelativistic Anisotropic Hydrodynamics}

\author{Aleksandr Kovalenko}
\affil{P.N. Lebedev Physical Institute, Moscow, Russia}

\maketitle

\begin{abstract}
Linear stability of a plane shock waves in ultrarelativistic anisotropic hydrodynamics is investigated. The properties of the amplitudes of perturbations of physical quantities are studied depending on the components of the wave vector of a small harmonic perturbation. Analytical calculations for the longitudinal and transverse propagation of shock wave normal with respect to the anisotropy axis (beam-axis) and numerical calculations for an arbitrary polar angle are carried out.
\end{abstract}

\section{Introduction}

The hydrodynamic approach is widely used to describe the evolution of matter created at the early stages of heavy ion collisions. Attempts to use dissipative hydrodynamic theories are presented in various papers \cite{Muronga:2002, Kolb:2003dz, Baier:2006um, Romatschke:2009im, Calzetta:2015}. However, the large pressure anisotropy, which appears at the early stages of heavy ion collisions due to the rapid longitudinal expansion, leads to a necessity of studing the effect of high-order gradients. Relativistic anisotropic hydrodynamics has been proposed as a theory where anisotropy is introduced explicitly as an appropriate parameter \cite{MartStr, RybFlor, Strickland, Alqahtani:2017mhy}. Anisotropic hydrodinamics produces solutions that are significally closer to the exact solutions of the Boltzmann equation than the standard viscous framework. This result was obtained both for longitudinally boost invariant and transversely homogeneous systems \cite{Florkowski:2013} and for Gubser flow \cite{Nopoush2015, Martinez2017}. It has been also shown that the anisotropic modeling is a promising approach in describing experimental data on heavy-ion collisions \cite{Mubarak:2017, Alqahtani:2018fcz}.

The formation of shock waves in a quark-guon medium during the heavy-ion collisions has been discussed for several decades \cite{Scheid:1974zz, Bouras:2009}. Mach cone generated by supersonic partons moving through the medium was stadied in the context of the jet-quenching phenomena \cite{Satarov:2005, Casalderrey-Solana_2007}. It has been shown that transverse shock waves in hot QCD matter can be produced by fluctuations of the local energy density (hot spots) and turbulence \cite{Gyulassy:1996ka, Gyulassy:1996br}. Appropriate description of stable shock waves in dissipative theories in general case is not possible. For the Israel-Stewart theory the existence of shock waves has been proved only for small Mach numbers \cite{Olson:1990, MajoranaMotta}. However, anisotropic relativistic hydrodynamics, even in the leading order, can give interesting solutions for shock waves.

In the framework of anisotropic hydrodynamics, it was found that a difference between the longitudinal and transverse pressures considered in the framework of anisotropic hydrodynamics leads to the anisotropy of sound propagation and the asymmetry of the Mach cone \cite{Kirakosyan:2018afm}. Previously, analytical expressions for the longitudinal and transverse propagation of shock wave normal with respect to the anisotropy axis (beam-axis) were obtained, and numerical calculations for an arbitrary polar angle were presented \cite{Kovalenko:2022}. The calculations were performed in the ultrarelativistic case with the assumption of constant anisotropy $\xi^{'} \simeq \xi$. Such effects as flow delfection and significant changes in the strength of shock waves depending on the parameters $\sigma = P^{'}/P, \ \xi$ and the polar angle were obtained. Some results lead to the question of the stability of shock waves against small perturbations of the discontinuity surface.

In present paper, the linear stability of shock waves in relativistic anisotropic hydrodynamics is investigated following the approach of \cite{Gardner1964, Russo1987}. The key point of this approach is the use of the Laplace transform for the amplitude of the perturbation of physical quantities. It should be noted that the result for the ultrarelativistic case was not discussed in the original work. The final equation is obtained using the law of conservation of particle number density, which no longer holds in the case of a massless gas. The plan of the paper is the following. The second section gives a presentation of this approach for the ultrarelativistic case, which construct the basis for an anisotropic description. The third chapter is devoted to the anisotropic case, where a brief presentation of the foundations of anisotropic relativistic hydrodynamics and a study of the stability of shock waves for the boundary cases of the location of the normal to the discontinuity surface is given. In fourth chapter the case of an arbitrary polar angle is considered.

\section{Isotropic case}

\subsection{Basic equations}

Consider an ultrarelativistic massless gas with the equation of state $\varepsilon = 3P$. In this case we do not assume the conservation law of the number of particles and concentrate only on the energy-momentum conservation law:
\begin{equation}
\partial_\mu T^{\mu\nu} = 0,
\label{equation}
\end{equation}
where
\begin{align}
T^{\mu\nu} &= (\varepsilon + P) U^\mu U^\nu - P g^{\mu\nu},
\label{tensor_iso}
\\
U^\mu &= (u_0, u_x, u_y, u_z), \ \ \ u_0 = \sqrt{1 + u_x^2 + u_z^2},
\end{align}
$P$ is pressure and $g^{\mu\nu}$ - the metric tensor.

We investigate a plane shock wave one-dimensional flow. Due to the isotropy, it is possible to fix any direction of the normal of the shock wave in space, and then, by transforming the coordinates, move to a system where one of the axes is directed along the normal. Therefore, without loss of generality, choose the normal vector $N^\mu = (0, 0, 0, 1)$. In this case, the discontinuity surface divides the space into two half-spaces $\Lambda_{+}$ for $z > 0$ and $\Lambda_{-}$ for $z < 0$. Since we are considering a one-dimensional flow, then one can put $v_x = v_y = 0$ and for the components of 4-velocity we have
\begin{align*}
u_0 = \frac{1}{\sqrt{1-v^2}}, \ \ \ u_z = \frac{v}{\sqrt{1-v^2}}.
\end{align*}
With a Lorentz transformation we can move into the rest frame of the shock wave. It is assumed that the direction of the flow is such that $v > 0$.

We consider small harmonic perturbation of the discontinuity surface that lead to the perturbed surface equation of the form
\begin{align}
f(t,x,y,z) = z - \eta e^{-i(\omega t + kx + ly)} = 0,
\end{align}
where $\eta$ is a small amplitude of the perturbation. We are interested in the mode of instability for which $k, l$ are real numbers and $\Image \omega > 0$. In this case, disturbance grows exponentially in time. Also we assume that in general case $l \neq 0, \ k \neq 0$. 

We expand the physical quantities to the first order
\begin{align}
u_0(t, x, y, z) &= u_0 + \delta u_0(t, x, y, z), \\
u_x(t, x, y, z) &= \delta u_x(t, x, y, z), \\
u_y(t, x, y, z) &= \delta u_y(t, x, y, z),\\
u_z(t, x, y, z) &= u_z +  \delta u_z(t, x, y, z),\\
P(t, x, y, z) &= P +  \delta P(t, x, y, z).
\end{align}
It is assumed that for $z \longrightarrow \pm \infty$ perturbations of physical quantities vanish so that $\delta u_0, \delta u_x, \delta u_y, \delta u_z,\delta P \rightarrow 0 $. Under such boundary conditions, exponential growth cannot be driven by energy transfer from distant boundaries. The region of the phase space (with the anisotropy parameter included) where this boundary condition is satisfied, together with $k, l \in \mathbb{R}, \Image \omega > 0$, forms the shock wave instability condition.

We introduce the vector $\textbf{W} = (\delta P,\delta u_x, \delta u_y, \delta u_z)$ for the perturbed quantities. Linearizing the equation (\ref{equation}) around a constant state, we obtain the following system of equations
\begin{equation}
A^\mu \partial_\mu \textbf{W} = 0.
\label{ur_iso}
\end{equation}
The matrices $A^\mu$ have the form
\begin{equation}
A^{0} = 
\begin{pmatrix}
-1 + (1+c_s^2)u_0^2 & 0 & 0 & 2(\varepsilon + P)u_0u_z \\
0 & (\varepsilon + P) u_0 &0 & 0 \\
0  & 0  & (\varepsilon + P) u_0 & 0  \\
(1+c_s^2)u_0 u_z & 0 & 0 & (\varepsilon + P) (u_0^2 + u_z^2)
\end{pmatrix}
\end{equation}

\begin{equation}
A^{1} = 
\begin{pmatrix}
0 & (\varepsilon + P) u_0 & 0 & 0 \\
1 & 0 &0 & 0 \\
0  & 0  & 0 & 0  \\
0 & (\varepsilon + P) u_z & 0 & 0
\end{pmatrix}, \ \ \ \ 
A^{2} = 
\begin{pmatrix}
0 & 0 & (\varepsilon + P) u_0 & 0 \\
0 & 0 &0 & 0 \\
1  & 0  & 0 & 0  \\
0 & 0 & (\varepsilon + P) u_z & 0
\end{pmatrix}
\end{equation}

\begin{equation}
A^{3} = 
\begin{pmatrix}
(1+c_s^2)u_0 u_z & 0 & 0 &  (\varepsilon + P) (u_0^2 + u_z^2) \\
0 & (\varepsilon + P) u_z &0 & 0 \\
0  & 0  & (\varepsilon + P) u_z & 0  \\
1 + (1+c_s^2)u_0^2 & 0 & 0 & 2(\varepsilon + P)u_0u_z
\end{pmatrix},
\end{equation}
whete $c_s^2 = (\partial P / \partial \varepsilon)_s$ is the speed of sound.

Since it is expected that the vector $\textbf{W}$ will inherit the perturbation of the discontinuity surface, we look for a solution of the specific form:
\begin{equation}
\textbf{W}(t, x, y, z) = \textbf{Y}(z) e^{-i(\omega t + kx + ly)},
\label{ansaz}
\end{equation}
where $\textbf{Y}(z)$ - amplitudes of perturbed quantities. By substituting (\ref{ansaz}) into the equations (\ref{ur_iso}) we have
\begin{equation}
\big( \omega A_0 + k A_1 + l A_2 + i \partial_z A_3 \big) \textbf{Y}(z) = 0.
\label{ww_iso_eq_1}
\end{equation}

It is assumed that $\textbf{Y}(z)$ admits the Laplace transform in the half-spaces $\Lambda_{-}: z < 0$ (behind the shock wave) and $\Lambda_{+}: z > 0$ (ahead the shock wave):
\begin{align*}
\widehat{\textbf{Y}}(q) &= \int_0^\infty e^{-qz} \textbf{Y}(z) dz \ \ \ \ \text{for} \ \ \Lambda_{+},
\\
\widehat{\textbf{Y}}(q) &= \int_0^\infty e^{-qz} \textbf{Y}(- z) dz \ \ \ \ \text{for} \ \ \Lambda_{-}.
\end{align*}
After the Laplace transform one finds
\begin{equation}
\big( \omega A_0 + k A_1 + l A_2 \pm i q A_3 \big) \widehat{\textbf{Y}}(q) \mp i A_3 \textbf{Y}(0)  = 0.
\label{matrix_eq_iso}
\end{equation}

Defining $q = \mp i m$ (for $\Lambda_\pm$) and $A = \omega A_0 + k A_1 + l A_2 + m A_3$, one can rewrite (\ref{matrix_eq_iso}) as
\begin{equation}
A \widehat{\textbf{Y}} (m) = \pm i A_3 \textbf{Y}(0), \ \ \text{in} \ \Lambda_\pm.
\label{matrix_eq_iso_2}
\end{equation}
Note that for the matrix $M = A_3^{-1} A$ the equation $\Det M = 0$ is an equation of the fourth degree in $m$ and is solvable in radicals. In particular,
\begin{equation}
\det M  = (m - m_0)(m-m_1)(m - m_2)(m - m_3).
\label{detM}
\end{equation}

For the system of equations
\begin{equation*}
A_3^{-1}A \widehat{\textbf{Y}} (m) = \pm i \textbf{Y}(0), \ \ \text{in} \ \Lambda_\pm,
\end{equation*}
with a known vector $\textbf{Y}(0)$, one can use Cramer's rule by introducing matrices $B_i$ $(i = 0, 1, 2, 3)$, that are constructed by replacing the $i$-th column of $M$ matrices, where the $i$-th column replaced by the $\textbf{Y}(0)$ column. Denoting $\det B_i = \Delta_i$, $\det M = \Delta_M$ we have following expression for the solution
\begin{equation*}
\widehat{\textbf{Y}} (m) = \Bigg(\frac{\Delta_0}{\Delta_M}, \frac{\Delta_1}{\Delta_M}, \frac{\Delta_2}{\Delta_M}, \frac{\Delta_3}{\Delta_M} \Bigg).
\end{equation*}
Using (\ref{detM}), we can rewrite the vector $\widehat{\textbf{Y}} (m)$ as a decomposition
\begin{equation}
\widehat{\textbf{Y}} (m) = \sum_i \textbf{C}_i \frac{1}{m - m_i},
\end{equation}
where $ \textbf{C}_i$ are some constant vectors. 

Then for the inverse Laplace transform we have
\begin{align}
\frac{1}{2i\pi} \int e^{-qz} \frac{1}{\pm iq - m_i} dq \thicksim e^{\mp i m_i z}, \ \ \text{in} \ \Lambda_\pm.
\label{exp_iso}
\end{align}
Therefore $\textbf{Y}(z)$ is the sum of plane waves $e^{i m_i z}$. This means that the system of differential equations (\ref{ww_iso_eq_1}) generates the linear system of equations, for which condition $\Det A = 0$ is necessary for their consistency. This condition provides characteristic equation for $m$ To understand the behavior of solutions $e^{\mp i m_i z}$, it is necessary to analyze the roots of this characteristic equation.

\subsection{The characteristic equation}

The equation $\Delta = \det A= 0$ reads as follows
\begin{equation}
\Omega^2\Big(\Omega^2 - c_s^2 [(k^2 + l^2)(1-v^2) + (m+v\omega)^2]\Big) = 0, \ \ \ \Omega = \omega + vm.
\label{char}
\end{equation}
Its solutions are the double root for $\Omega = 0$ and the two roots of the quadratic equation:
\begin{align}
m_0 &= m_1 = -\frac{\omega}{v}, \\
m_2 &= \frac{-v\omega(1-c_s^2) + c_s\sqrt{1-v^2}\sqrt{(k^2 + l^2)(v^2-c_s^2) + \omega^2(1-v^2)}}{v^2-c_s^2}, \\
m_3 &= \frac{-v\omega(1-c_s^2) - c_s\sqrt{1-v^2}\sqrt{(k^2 + l^2)(v^2-c_s^2) + \omega^2(1-v^2)}}{v^2-c_s^2}.
\label{m3}
\end{align}
The root $m_{0,1} = -\omega/v$ corresponds to the entropy-vortex perturbations propagating with the gas \cite{Russo1987}. We assume that $\Real \omega > 0$. Then for the double root $m_{0, 1}$ we have
\begin{align}
\Real \omega \geqslant 0 , \ \Image \omega > 0 \Leftrightarrow \Real m_{0, 1} \leqslant 0 , \ \Image m_{0, 1} < 0
\label{conditions0}
\end{align}
behind and ahead the shock wave.

To analyze the next two roots, we introduce the following definition $q = u_0^{2}(k^2 + l^2)(v^2-c_s^2)$, obtaining
\begin{align}
m_{2, 3} &= \frac{-v u_0^2\omega(1-c_s^2) \pm c_s\sqrt{q + \omega^2}}{u_0^2(v^2-c_s^2)}.
\label{m23}
\end{align}
Let $\omega = \omega_R + i \omega_I$, then $\sqrt{q + \omega^2} = \sqrt{z_R + i z_I}$,
where
\begin{align*}
z_R = q + \omega_R^2 - \omega_I^2, \ \ \ z_I = 2\omega_R\omega_I.
\end{align*}
It is known that for the square root of a complex number is calculated one can write
\begin{align*}
\sqrt{z_R + i z_I} = \pm \Bigg[ \sqrt{\frac{\sqrt{z_R^2 + z_I^2} + z_R}{2}} + i \sign(z_I)   \sqrt{\frac{\sqrt{z_R^2 + z_I^2} - z_R}{2}}\Bigg].
\end{align*}
Since the choice of the $\pm$ sign before the brackets only swaps the roots $m_2$ and $m_3$, we can restrict ourselves to considering the positive sign. We can write expressions for the real and imaginary parts of $m_2$ with $\sign(z_I) = \sign(\omega_R \omega_I) = 1$ as follows
\begin{align*}
u_0^2(v^2-c_s^2)\Real m_2 &= -\omega_R \Bigg(v u_0^2 (1-c_s^2) - \frac{c_s}{\omega_R} \sqrt{\frac{\sqrt{z_R^2 + z_I^2} + z_R}{2}}\Bigg),\\
u_0^2(v^2-c_s^2)\Image m_2 &= -\omega_I \Bigg(v u_0^2 (1-c_s^2) - \frac{c_s}{\omega_I} \sqrt{\frac{\sqrt{z_R^2 + z_I^2} - z_R}{2}}\Bigg).
\end{align*}
In the half-space $\Lambda_{-}$ behind the shock wave $v^2 - c_s^2 > 0$ and $q > 0$, so that
\begin{align*}
&\frac{1}{ |\omega_I|} \sqrt{\frac{\sqrt{z_R^2 + z_I^2} - z_R}{2}} < \frac{1}{ |\omega_I|} \sqrt{\frac{\sqrt{( \omega_R^2 + \omega_I^2)^2} - (\omega_R^2 - \omega_I^2)}{2}} = 1,\\
&\frac{1}{ \omega_R} \sqrt{\frac{\sqrt{z_R^2 + z_I^2} + z_R}{2}} < \frac{1}{ \omega_R} \sqrt{\frac{\sqrt{( \omega_R^2 + \omega_I^2)^2} + (\omega_R^2 - \omega_I^2)}{2}} = 1,
\end{align*}
for which it follows that
\begin{align}
\Real \omega  \geqslant 0 , \ \Image \omega > 0 &\Leftrightarrow \Real m_{2}  \leqslant 0 , \ \Image m_{2} < 0 \ \ \text{in} \ \Lambda_{-}.
\end{align}

For the half-space $\Lambda_{+}$ we have $v^2 - c_s^2 < 0$ and $q < 0$, so one obtains
\begin{align*}
&\frac{1}{ |\omega_I|} \sqrt{\frac{\sqrt{z_R^2 + z_I^2} - z_R}{2}} > 1, \ \ \ \ \frac{1}{ \omega_R} \sqrt{\frac{\sqrt{z_R^2 + z_I^2} + z_R}{2}} > 1.
\end{align*}
Therefore in $\Lambda_{+}$ we have similar inequalities
\begin{align*}
\Real \omega  \geqslant0 , \ \Image \omega > 0 &\Leftrightarrow \Real m_{2}  \leqslant 0 , \ \Image m_{2} < 0 \ \ \text{in} \ \Lambda_{+}.
\end{align*}

Expressions for the real and imaginary parts of the root $m_3$ read
\begin{align*}
u_0^2(v^2 - c_s^2)\Image m_3 &= -\omega_I \Bigg(v u_0^2 (1-c_s^2) + \frac{c_s}{\omega_I} \sqrt{\frac{\sqrt{z_R^2 + z_I^2} - z_R}{2}}\Bigg), \\
u_0^2(v^2 - c_s^2)\Real m_3 &= -\omega_R \Bigg(v u_0^2 (1-c_s^2) + \frac{c_s}{ \omega_R} \sqrt{\frac{\sqrt{z_R^2 + z_I^2} + z_R}{2}}\Bigg).
\end{align*}
Since the sign before the square root is positive here, it is obvious that
\begin{align}
\Real \omega  \geqslant0 , \ \Image \omega > 0 &\Leftrightarrow \Real m_{3}  \leqslant 0 , \ \Image m_{3} < 0 \ \ \text{in} \ \Lambda_{-},\\
\Real \omega  \geqslant 0 , \ \Image \omega > 0 &\Leftrightarrow \Real m_{3}  \geqslant 0 , \ \Image m_{3} > 0 \ \ \text{in} \ \Lambda_{+}.
\label{conditions3}
\end{align}

\subsection{Solution for $\widehat{\textbf{Y}} (m)$}

Now we can specify the behavior of $\textbf{Y}(x)$ in the $\Lambda_\pm$ half-spaces. For $\Lambda_{-} \ (z < 0)$ we obtain $\Image m_0 < 0, \ \Image m_2 < 0, \Image m_3 < 0$, then for $z \longrightarrow \infty$ we have exponential growth of (\ref{exp_iso}) that violates boundary condition, hence we need to put $\textbf{C}_{1},\textbf{C}_{2}, \textbf{C}_{3} = 0$. For half-space $\Lambda_{+} \ (z > 0)$ the inequalities are $\Image m_0 < 0, \ \Image m_2 < 0, \Image m_3 > 0$, that leads to the condition $\textbf{C}_{3} = 0$.

Now, we need an expression for the vector $\textbf{Y}(0)$. This can be obtained using the matching equation for the energy-momentum tensor at the discontinuity surface \cite{landau2013course, Mitchell}. Let $N^\mu$ be the normal to the discontinuity surface, then
\begin{equation}
N_\mu T^{\mu\nu} = N_\mu T^{'\mu\nu},
\label{matching_iso}
\end{equation}
where the prime denotes variables in the half-space $\Lambda_+$ (behind the shock wave). For the unperturbed case with $N^\mu = (0,0,0,1)$ we have two equations
\begin{align}
(\varepsilon + P)u_0 u_z &= (\varepsilon^{'} + P^{'})u_0^{'} u_z^{'}, 
\label{normal_eq_iso_1}
\\
(\varepsilon + P)u_z^2 - P &= (\varepsilon^{'} + P^{'})u_z^{'2} - P^{'}.
\label{normal_eq_iso_2}
\end{align}

The expression for the normal of the perturbed surface obtained from the equation
\begin{equation*}
f(t, x, y, z) = z - \eta e^{-i(\omega t + kx + ly)} = 0
\end{equation*}
leads to
\begin{equation}
N_\mu = \partial_\mu f = (i \eta \omega, i\eta k, i\eta l, 1).
\label{normal_iso}
\end{equation}
Here and below, we will neglect the exponent, meaning that it is included in the amplitude $\eta$. Since we found that $\textbf{Y}(0) = 0$ for the half-space $\Lambda_{-}$, then for the upstream and downstream 4-velocity vectors we have
\begin{align}
U^\mu &= (u_0, 0, 0, u_z), 
\label{upstream_iso}
\\
U^{'\mu} &= (u_0^{'} + \delta u_0^{'}, \delta u_x^{'}, \delta u_y^{'}, u_z^{'} + \delta u_z^{'}).
\label{downstream_iso}
\end{align}
It follows from $U^{'\mu} U_\mu^{'} = 1$ that $u^{'}_0 \delta u^{'}_0 = u^{'}_z \delta u ^{'}_z$. Substituting energy-momentum tensor with perturbed vectors (\ref{upstream_iso} - \ref{downstream_iso}) and the normal (\ref{normal_iso}) into the equations (\ref{matching_iso}) and performing some transformations using the equations (\ref{normal_eq_iso_1} - \ref{normal_eq_iso_2}), we obtain the following components of $\textbf{Y}(0)$
\begin{align}
Y^0(0) &= \delta P^{'}  = -2 i  \eta \omega  (\varepsilon^{'} + P^{'}) c_s^2 \frac{(u_0 u_z^{'} - u_0^{'} u_z)(u_0 u_0^{'} - u_z u_z^{'})}{(c_s^2-(1-c_s^2)u_z^{'2})u_z u_0}\ u_z^{'} u_0^{'},
\\
Y^1(0) &=\delta u_x^{'} = i \eta k \frac{u_0 u_z^{'} - u_0^{'} u_z}{u_0}, 
\\
Y^2(0) &=\delta u_y^{'} = i \eta l \frac{u_0 u_z^{'} - u_0^{'} u_z}{u_0}, 
\\
Y^3(0) &= \delta u_z^{'} = i  \eta \omega \frac{(u_0 u_z^{'} - u_0^{'} u_z)(u_0 u_0^{'} - u_z u_z^{'})}{(c_s^2-(1-c_s^2)u_z^{'2})u_z u_0}\ (u_z^{'2} + c_s^2(1+u_z^{'2}))u_0^{'}.
\end{align}

Now it is possible to solve the equation
\begin{equation*}
M \widehat{\textbf{Y}} (m) = i  \textbf{Y}(0)
\end{equation*}
using Cramer's rule as described above. Note that $\Delta_i$ are a polynomial of the fourth degree in $m$, furthermore
\begin{align}
\Delta_0 &= (m -m_0)^2 P^{(1)}_0(m),
\label{poly0}
\\
\Delta_1 &= (m -m_0) P^{(2)}_1(m),\\
\Delta_2 &= (m -m_0) P^{(2)}_2(m),\\
\Delta_3 &= (m -m_0) P^{(2)}_3(m),
\label{poly3}
\end{align}
where $ P^{(n)}_i(m)$ is a polynomial of the $n$th degree in $m$. Since $\Delta_M = (m - m_0)^2(m - m_2)(m - m_3)$, we can write the following decomposition
\begin{equation}
\widehat{\textbf{Y}} (m) = \frac{\textbf{C}_{1}}{m - m_0} + \frac{\textbf{C}_{2}}{m - m_2} + \frac{\textbf{C}_{3}}{m - m_3}.
\label{decom}
\end{equation}
As already mentioned we are interested in the condition $\textbf{C}_{3} = 0$. One can find expressions for the components of the vector $\textbf{C}_3$, using formulas (\ref{poly0} - \ref{poly3}) and decomposition (\ref{decom}):
\begin{align}
C_{30} &= -R \frac{\eta c_s (\varepsilon + P)(v^{'} - v)}{2vv^{'2}(v^{'2} - c_s^2)(1-v^{'2})^2Q},
\\
C_{31} &=  -R \frac{\eta k (v^{'} - v)}{2vv^{'}(1-v^{'2})^{3/2}Q [c_s\omega + v^{'} Q]},
\\
C_{32} &= -R \frac{\eta l (v^{'} - v)}{2vv^{'}(1-v^{'2})^{3/2}Q [c_s\omega + v^{'} Q]},
\\
C_{33} &=  -R \frac{\eta(Qc_s + \omega v^{'}) (v^{'} - v)}{2vv^{'2}(v^{'2} - c_s^2) (1-v^{'2})^2 Q  [c_s\omega + v^{'} Q]},
\end{align}
where
\begin{align}
Q &= \sqrt{(k^2 + l^2)(v^{'2}-c_s^2)/(1-v^{'2}) + \omega^2},
\\
R &= 2v^{'}c_s(1-vv^{'})\omega Q + (v^{'2} + c_s^2)(1-vv^{'})\omega^2 + vv^{'} (v^{'2} - c_s^2)(k^2+l^2).
\end{align}

It is seen that the common factor for the components of the vector $\textbf{C}_{3}$ is $R$, hence the obvious requirement for $\textbf{C}_{3} = 0$ is the condition $R = 0$.

\subsection{Unstable Mode}

We will assume an ideal equation of state $\varepsilon = 3P$ which is appropriate for a massless gas. It is known that the speed of sound in such a medium is $c_s^2 = 1/3$. Moreover, the shock wave solution leads to the relation $vv^{'} = 1/3$. Formulae allow us to work with only one quantity - the downstream velocity $v^{'}$. Also one can conclude from the characteristic equation (\ref{char}) and the expression for $m_3$ (\ref{m3}) that
\begin{align}
k^2 + l^2 &= \frac{(mv^{'} + \omega)^2 - c_s^2(m+v^{'}\omega)^2}{c_s^2(1-v^{'2})},
\\
Q &= \frac{c_s^2(m+v^{'}\omega) - v^{'}(mv^{'} + \omega)}{c_s(1-v^{'2})}.
\end{align}
Therefore, in the ultrarelativistic case the condition $R = 0$ is equivalent to the equation
\begin{align}
\varphi^2 - 2v^{'}\varphi - (1-v^{'2}) = 0,
\label{cond}
\end{align}
where $\varphi = \Omega /m$, $\Omega = \omega + vm$. This equation has only real solutions
\begin{align}
\varphi_{1,2} = v^{'} \pm 1.
\end{align}

We have $\Real \varphi \geqslant v^{'}, \ \Image \varphi < 0$, since we are considering a mode in which $\Image m_3 > 0, \Real m_3 \geqslant 0$. However, this does not entirely specify the range of the variable $\varphi$. The domain of $\varphi$ can be determined using equation (\ref{char}) (see Appendix A). For real values of $\varphi$ we find that the following inequalities should hold
\begin{align}
v^{'} \leqslant \varphi \leqslant \frac{c_s(1- v^{'2})}{(1-c_sv^{'})}.
\end{align}

It is seen that both solutions do not fall into this area. This means that the mode of instability does not exist for the ultrarelativistic case.

\section{Anisotropic case}

\subsection{Anisotropic relativistic hydrodynamics}

The framework of anisotropic hydrodynamics we use in this paper is based on the kinetic theory approach  \cite{MartStr,StrRom1,StrRom2}, where  one assumes that the distribution function $f$ is a ansatz of Romatschke-Strickland form
\begin{equation}
f(x,p) = f_{\textrm{iso}}\Bigg( \frac{\sqrt{p^\mu \Xi_{\mu\nu}(x) p^\nu}}{\Lambda(x)}\Bigg),
\label{Kin1}
\end{equation} 
where $\Lambda(x)$ is a coordinate-dependent temperature-like momentum scale and $\Xi_{\mu\nu}(x)$ quantifies coordinate-dependent momentum anisotropy. In what follows we consider one-dimensional anisotropy such that $ p^\mu \Xi_{\mu\nu}p^\nu = \mathbf{p}^2 +\xi(x) p_\parallel^2$  in the local rest frame (LRF).

To construct the energy-momentum tensor as the second moment of the distribution function we define a general orthogonal tensor basis $U^\mu, X^\mu, Y^\mu, Z^\mu$ which in the LRF reads
\begin{align}
U^\mu_{LRF} &= (1,0,0,0),\\
X^\mu_{LRF} &= (0,1,0,0),\\
Y^\mu_{LRF} &= (0,0,1,0),\\
Z^\mu_{LRF} &= (0,0,0,1).
\end{align}
Since we consider one-dimensional (longitudinal) anisotropy, one can write the energy-momentum tensor $T^{\mu\nu}$ in terms of four-velocity vector $U^\mu$ and space-like longitudinal vector $Z^\mu$ as follows \cite{Martinez2012}
\begin{equation}
T^{\mu\nu} = (\varepsilon + P_\perp)U^\mu U^\nu - P_\perp g^{\mu\nu} + (P_\parallel - P_\perp)Z^\mu Z^\nu,
\label{T_true}
\end{equation}
where $P_\parallel$ and  $P_\perp$ is longitudinal (towards anisotropy direction) and transverse pressure respectively. In the LRF the expression \eqref{T_true} takes the form
\begin{equation*}
T^{\mu\nu} = {\rm diag} (\varepsilon,P_\perp,P_\perp,P_\parallel).
\end{equation*}
In the case of massless gas the condition of the tracelessness of the energy-momentum tensor leads to the relation $\varepsilon = 2P_\perp + P_\parallel$.

It is convenient to rewrite the four-vector $U^\mu (x)$ in terms of the longitudinal rapidity $\vartheta (x)$, the time-like velocity $u_0 = \sqrt{1+u_x^2+u_y^2}$ and transverse velocities $u_x, u_y$ \cite{Ryblewski2011}
\begin{equation}
U^\mu = (u_0 \cosh \vartheta, u_x, u_y, u_0 \sinh \vartheta).
\label{u_aniso}
\end{equation}
Then vector $Z^\mu$ takes the form
\begin{equation}
Z^\mu = (\sinh \vartheta, 0,0, \cosh \vartheta).
\label{z_aniso}
\end{equation}

It is important to note that the dependence on the anisotropy parameter $\xi$ can be factorized \cite{MartStr}:
\begin{align}
\varepsilon &= \int \frac{d^3 p}{(2\pi)^3} p^0 f_{\textrm{iso}}\Bigg(\frac{\sqrt{\textbf{p}^2 + \xi(x) p_\parallel^2}}{\Lambda(x)}\Bigg) = R(\xi) \varepsilon_{\textrm{iso}} (\Lambda),
\label{e} \\
P_\perp &= \int \frac{d^3 p}{(2\pi)^3} \frac{p_\perp^2}{2 p_0}f_{\textrm{iso}}\Bigg(\frac{\sqrt{\textbf{p}^2 + \xi(x) p_\parallel^2}}{\Lambda(x)}\Bigg) = R_\perp (\xi) P_{\textrm{iso}} (\Lambda),
\label{pT} \\
P_\parallel &= \int \frac{d^3 p}{(2\pi)^3} \frac{p_\parallel^2}{p_0}f_{\textrm{iso}}\Bigg(\frac{\sqrt{\textbf{p}^2 + \xi(x) p_\parallel^2}}{\Lambda(x)}\Bigg) = R_\parallel (\xi) P_{\textrm{iso}} (\Lambda),
\label{pL}
\end{align}
where the anisotropy-dependent factors $R_\perp(\xi)$ and $R_\parallel(\xi)$ are \cite{MartStr}
\begin{equation}
\label{RTL}   
R_\perp(\xi) = \frac{3}{2\xi} \Bigg( \frac{1 + (\xi^2-1)R(\xi)}{1+\xi}\Bigg), \;\;\;\;\;\;
R_\parallel(\xi) = \frac{3}{\xi} \Bigg( \frac{(\xi+1)R(\xi) -1}{1+\xi}\Bigg), 
\end{equation}
where, in turn, 
\begin{equation}
\label{R}
  R(\xi) = \frac{1}{2} \Bigg( \frac{1}{1+\xi} + \frac{\arctan \sqrt{\xi}}{\sqrt{\xi}}\Bigg). 
\end{equation}
The ultrarelativistic condition $\varepsilon_{\textrm{iso}} = 3  P_{\textrm{iso}}$ leads to the following relation between the anisotropic functions:
\begin{equation}
\label{RT+RP}
2 R_\perp(\xi) + R_\parallel(\xi) = 3 R(\xi).
\end{equation}

In the preceding paper \cite{Kirakosyan:2018afm} we have derived the following equation describing propagation of sound in relativistic anisotropic hydrodynamics with longitudinal anisotropy:
\begin{equation}
\partial^2_t \;  n^{(1)} = \left ( c^2_{s \perp} \; \partial^2_\perp + c^2_{s \parallel} \; \partial^2_z \right)  n^{(1)},
\end{equation}
where $n^{(1)}$ is a (small) density fluctuation and $c_{s \perp}$ and $c_{s  \parallel}$ stand for anisotropy-dependent transverse and longitudinal speed of sound respectively. The explicit expressions for $c^2_{s \perp}$ and $c^2_{s \parallel}$ read
\begin{equation}
\label{csTR2}
c^2_{s \perp}  =  \frac{R_\perp}{3 R}, \;\;\; c^2_{s \parallel} = \frac{R_\parallel}{3 R}.
\end{equation}

Before introducing a perturbation on the discontinuity surface, we consider the process of linearization of the equation $\partial_\mu T^{\mu\nu}$. We assume that the anisotropy is constant $\xi = \xi^{'}$, so we perturb the isotropic quantities. The isotropic pressure $P_{\textrm{iso}}$, longitudinal rapidity $\vartheta$ and four-velocity components $u_x, \ u_y$ are linearized around constant state as
\begin{align}
P_{\textrm{iso}}(t, x, y, z) &= P + \delta P(t, x, y, z), 
\label{r1}
\\
\vartheta(t, x, y, z) &= \vartheta+  \delta\vartheta(t, x, y, z), 
\label{r2} \\
u_x(t, x, y, z) &=  u_x + \delta u_x(t, x, y, z), \\
u_y(t, x, y, z) &= u_y + \delta u_y(t, x, y, z).
\label{r3}
\end{align}

From the equations $\partial_\mu T^{\mu\nu} = 0$ we obtain
\begin{equation}
A_{\textrm{aniso}}^\mu \partial_\mu \textbf{W}_{\textrm{aniso}} = 0,
\label{eq_aniso}
\end{equation}
where
\begin{equation}
\textbf{W}_{\textrm{aniso}} = (\delta P, \delta u_x, \delta u_y, \delta \vartheta),
\label{ww_aniso}
\end{equation} 
and the matrices $A_{\textrm{aniso}}^\mu$ are
\begin{equation}
A_{\textrm{aniso}}^{0} = 
\begin{pmatrix}
R_\perp + R_1 u_0^2 \cosh^2\vartheta + R_3 \sinh^2\vartheta& 2 PR_1 u_j\cosh^2\vartheta & P (2  R_2 + R_1u_0^2) \sinh(2 \vartheta) \\
R_1 u_i u_0  \cosh\vartheta & P R_1 (u_0^2\delta_{ij} + u_i u_j) \cosh\vartheta / u_0 & P R_1 u_i u_0 \sinh\vartheta \\
P (2  R_2 + R_1u_0^2) \sinh(2 \vartheta)/2 & PR_1 u_j \sinh(2\vartheta) & P (2  R_2 + R_1u_0^2) \cosh(2 \vartheta)
\end{pmatrix},
\label{matrix_aniso_0}
\end{equation}

\begin{equation}
A_{\textrm{aniso}}^{k} = 
\begin{pmatrix}
R_1 u_k u_0 \cosh\vartheta & P R_1 (u_0^2\delta_{kj} + u_k u_j) \cosh\vartheta/ u_0 & P R_1 u_k u_0 \sinh\vartheta \\
R_\perp \delta_{ik} + R_1 u_k u_i & P R_1 (u_k \delta_{ij} + u_i \delta_{jk} ) & 0 \\
P R_1 u_k u_0 \sinh\vartheta & P R_1 (u_0^2\delta_{kj} + u_k u_j) \sinh\vartheta/ u_0 & P R_1 u_k u_0 \cosh\vartheta
\end{pmatrix},
\end{equation}

\begin{equation}
A_{\textrm{aniso}}^{3}  = 
\begin{pmatrix}
P (2  R_2 + R_1u_0^2) \sinh(2 \vartheta)/2 & PR_1 u_j \sinh(2\vartheta) & P (2  R_2 + R_1u_0^2) \cosh(2 \vartheta) \\
R_1 u_i u_0  \sinh\vartheta & P R_1 (u_0^2\delta_{ij} + u_i u_j) \sinh\vartheta / u_0 & P R_1 u_i u_0 \cosh\vartheta \\
R_\perp + R_2 \cosh(2 \vartheta) + R_1 u_0^2\sinh^2\vartheta & 2PR_1u_j \sinh^2\vartheta & P (2  R_2 + R_1u_0^2) \sinh(2 \vartheta)
\end{pmatrix},
\label{matrix_aniso_3}
\end{equation}
where, in turn,
\begin{align}
R_1 &= (R_\parallel + 3 R_\perp), \\
R_2 &= (R_\parallel + R_\perp), \\ 
R_3 &= (R_\parallel - R_\perp).
\end{align}
The indices $i ,j$ correspond to rows and columns and $k = 1, 2$. For velocities we have $u_1 = u_x, u_2 = u_y$. 

\subsection{Stability of the longitudinal normal shock wave}

The normal to the undisturbed discontinuity surface is directed along the anisotropy direction, i.e. $N^\mu = (0, 0, 0, 1)$. We will assume that $u_x = u_y = 0$ and define the same small harmonic perturbation of the discontinuity surface as in the isotropic case: 
\begin{align}
f(t,x,y,z) = z - \eta e^{-i(\omega t + kx + ly)} = 0.
\end{align}

In terms of velocities one finds
\begin{align*}
\sinh\vartheta^{'} = \frac{v^{'}}{\sqrt{1-v^{'2}}}, \ \cosh\vartheta^{'} =  \frac{1}{\sqrt{1-v^{'2}}}.
\end{align*}

It is assumed that a solution for the vector $\textbf{W}_{\textrm{aniso}} = \textbf{W}_\parallel$ in equation (\ref{ww_aniso}) has the form 
\begin{align}
\textbf{W}_\parallel(t, x, y, z) = \textbf{Y}_\parallel(z) e^{-i(\omega t + kx +ly)}.
\end{align}
We denote matrices $A^{i}_{\textrm{aniso}}$ at $u_x = u_y = 0$ as the matrices $A^{i}_{\parallel}$. Laplace transformation for the amplitude vector $\textbf{Y}_\parallel$ gives the following equation
\begin{equation}
A_\parallel \widehat{\textbf{Y}}_\parallel (m) = \pm i A_{\parallel 3} \textbf{Y}_\parallel(0),
\label{eq_parallel_1}
\end{equation}
in half-spaces $\Lambda_{\pm}$ respectively, where $A_\parallel = \omega A_{\parallel 0} + k A_{\parallel 1} + l A_{\parallel 2} + m A_{\parallel 3}$.

Substitution of the ansatz (\ref{ansaz}) into the equations (\ref{eq_aniso}) leads to a characteristic equation $\Det A_\parallel = 0$, which reads
\begin{equation}
\Omega^2\Big(2\Omega^2 - [(1 - c_{s\parallel}^2)(k^2 + l^2)(1-v^2) + 2c_{s\parallel}^2(m+v\omega)^2]\Big) = 0, \ \ \ \Omega = \omega + vm,
\label{char_parallel}
\end{equation}
where 
\begin{equation}
c_{s\parallel} = \frac{R_\parallel}{3R}
\end{equation}
is the longitudinal speed of sound.

Solving the equation for $m$, we obtain four roots
\begin{align}
m_{0, 1} &= - \frac{\omega}{v},
\label{m0_parallel}
\\
m_2 &= \frac{1 - v^2}{4(v^2-c_{s\parallel}^2)}\Bigg( -4v\omega(1-c_{s\parallel}^2) + 2\sqrt{2}\sqrt{2c_{s\parallel}^2\omega^2 + \frac{(1-c_{s\parallel}^2)(k^2+l^2)(v^2-c_{s\parallel}^2)}{1-v^2}} \Bigg),
\\
m_3 &= \frac{1 - v^2}{4(v^2-c_{s\parallel}^2)}\Bigg( -4v\omega(1-c_{s\parallel}^2) - 2\sqrt{2}\sqrt{2c_{s\parallel}^2\omega^2 + \frac{(1-c_{s\parallel}^2)(k^2+l^2)(v^2-c_{s\parallel}^2)}{1-v^2}} \Bigg).
\label{m3_parallel}
\end{align}

Root analysis is carried out in a similar way to the isotopic case and leads to the same relations:
\begin{align}
\Real \omega  \geqslant0 , \ \Image \omega > 0 &\Leftrightarrow \Real m_{0, 1}  \leqslant 0 , \ \Image m_{0, 1} < 0 \ \ \text{in} \ \Lambda_{\pm},\\
\label{conditions0_parallel}
\Real \omega  \geqslant 0 , \ \Image \omega > 0 &\Leftrightarrow \Real m_{2}  \leqslant 0 , \ \Image m_{2} < 0 \ \ \text{in} \ \Lambda_{-},\\
\Real \omega  \geqslant0 , \ \Image \omega > 0 &\Leftrightarrow \Real m_{2}  \leqslant 0 , \ \Image m_{2} < 0 \ \ \text{in} \ \Lambda_{+},\\
\Real \omega  \geqslant0 , \ \Image \omega > 0 &\Leftrightarrow \Real m_{3}  \leqslant 0 , \ \Image m_{3} < 0 \ \ \text{in} \ \Lambda_{-},\\
\Real \omega  \geqslant 0 , \ \Image \omega > 0 &\Leftrightarrow \Real m_{3}  \geqslant 0 , \ \Image m_{3} > 0 \ \ \text{in} \ \Lambda_{+}.
\label{conditions_parallel}
\end{align}

Since the longitudinal case is technically the same as the above-considered isotropic case, we have the same decomposition (\ref{decom}) for the vector $\textbf{Y}_\parallel(z)$. To satisfy the boundary condition $\delta u_x, \delta u_y, \delta \vartheta,\delta P \rightarrow 0 $ at $z \longrightarrow \pm \infty$ we should put again $\textbf{C}_{i} = 0, i = 1,2,3$ in $\Lambda_{-}$ and $\textbf{C}_{3 } = 0$ in $\Lambda_{+}$.

The matching equation for the energy-momentum tensor on the discontinuity surface $N_\mu T^{\mu\nu} = N_\mu T^{'\mu\nu}$ in the unperturbed case leads to the equations
\begin{align}
(R_\perp - R_\parallel)(P\sinh(2\vartheta) - P^{'}\sinh(2\vartheta^{'})) &= 0, \\
(P - P^{'})R_\perp - P(R_\perp + R_\parallel)\cosh(2\vartheta) + P^{'}(R_\perp + R_\parallel)\cosh(2\vartheta^{'}) &= 0.
\label{matching_eqs_parallel}
\end{align}
from which in terms of velocities one obtains (see \cite{Kovalenko:2022})
\begin{align}
v v^{'} = \frac{R_\parallel}{3R}.
\end{align}

Since we found that $\textbf{Y}_\parallel(0) = 0$ in half-space $\Lambda_{-}$ then for the velocities we obtain
\begin{align}
U^\mu &= (\cosh\vartheta, 0, 0, \sinh\vartheta),
\label{perturbed_parallel_1} \\
Z^\mu &= (\sinh\vartheta, 0, 0, \cosh\vartheta), \\
U^{'\mu} &= (\cosh\vartheta^{'} + \delta\vartheta^{'}\sinh\vartheta^{'}, \delta u_x^{'}, \delta u_y^{'}, \sinh\vartheta^{'} + \delta\vartheta^{'}\cosh\vartheta^{'}),\\
Z^{'\mu} &= (\sinh\vartheta^{'} + \delta\vartheta^{'}\cosh\vartheta^{'}, 0, 0, \cosh\vartheta^{'} + \delta\vartheta^{'}\sinh\vartheta^{'}).
\label{perturbed_parallel_4}
\end{align}

Sabstituting the perturbed vectors (\ref{perturbed_parallel_1} - \ref{perturbed_parallel_4}) into the energy-momentum tensor and perturbed normal vector $N_\mu = (i \eta \omega, i\eta k, i\eta l, 1)$ into the matching condition one finds
\begin{align}
Y^0_\parallel(0) &= \delta P^{'}  = 8 i \eta \omega P^{'} \frac{R_\perp R_2 \sinh\vartheta^{'} \cosh\vartheta^{'} }{3 R R_\parallel },
\\
Y^1_\parallel(0) &=\delta u_x^{'} = -2 i \eta k \frac{R_\perp R_2 (R_\parallel - 2 R_\perp \sinh^2\vartheta^{'}) }{3 R R_\parallel R_1 \sinh\vartheta^{'}},
\\
Y^2_\parallel(0) &=\delta u_y^{'} = -2 i \eta l \frac{R_\perp R_2 (R_\parallel - 2 R_\perp \sinh^2\vartheta^{'}) }{3 R R_\parallel R_1 \sinh\vartheta^{'}},
\\
Y^3_\parallel(0) &= \delta \vartheta^{'} = -2 i \eta \omega \frac{[R_\parallel + 2R_2 \sinh^2\vartheta^{'} ]}{3 R R_\parallel}.
\end{align}

Solving the equation (\ref{eq_parallel_1}) using Cramer's rule, we find the following expressions for the components of the vector $\textbf{C}_3$
\begin{align}
C_{30} &= -R \eta P^{'} \frac{ 1 - c_{s\parallel}^4}{2v^{'2}c_{s\parallel}^2(1-v^{'2})Q},
\\
C_{31} &= -R \eta k \frac{(1 - c_{s\parallel}^4)(1 - c_{s\parallel}^2)(v^{'2}-c_{s\parallel}^2)}{c_{s\parallel}^2 (3-c_{s\parallel}^2)v^{'}(1-v^{'2})^{3/2} Q [2c_{s\parallel}^2\omega + v^{'}  Q]},
\\
C_{32} &= -R \eta l  \frac{(1 - c_{s\parallel}^4)(1 - c_{s\parallel}^2)(v^{'2}-c_{s\parallel}^2)}{c_{s\parallel}^2 (3-c_{s\parallel}^2)v^{'}(1-v^{'2})^{3/2} Q [2c_{s\parallel}^2\omega + v^{'}  Q]},
\\
C_{33} &=   R \eta \frac{(1 - c_{s\parallel}^2)(Q + 2v^{'}\omega)}{2v^{'2}c_{s\parallel}^2 (1-v^{'2}) Q [2c_{s\parallel}^2\omega + v^{'}  Q]},
\end{align}
where
\begin{align}
Q &= \sqrt{2(k^2 + l^2)(v^{'2}-c_{s\parallel}^2)(1-c_{s\parallel}^2)/(1-v^{'2}) + 4\omega^2c_{s\parallel}^2},
\\
R &= 2v^{'}\omega Q + 2 (v^{'2} + c_{s\parallel}^2)\omega^2 + (v^{'2} - c_{s\parallel}^2) (k^2 + l^2).
\end{align}

Together with (\ref{char_parallel}), (\ref{m3_parallel}) one finds that the condition $R = 0$ is equivalent to the equation
\begin{align}
\varphi^2 - 2v^{'}\varphi - (1-v^{'2}) = 0,
\label{cond_parallel}
\end{align}
where $\varphi = \Omega /m$, $\Omega = \omega + vm$. The equation (\ref{cond_parallel}) is identical with the on eobtained in the isotropic case. Therefore, we have the same roots $v^{'} \pm 1$. We also obtain from their characteristic equation that for the real $\varphi$ the following inequalities must be satisfied
\begin{align}
v^{'} \leqslant \varphi \leqslant \frac{c_s(1- v^{'2})}{(1-c_sv^{'})},
\end{align}
which leads to the conclusion that of the mode of instability that we looked for is absent.

\subsection{Stability of the transverse normal shock wave}

The normal to the undisturbed discontinuity surface orthogonal to the anisotropy direction, i.e. $N^\mu = (0, 1, 0, 0)$. Now, we assume that $u_z = \vartheta = 0$ and introduce the small harmonic perturbation
\begin{align}
f(t,x,y,z) = x - \eta e^{-i(\omega t + ly + mz)} = 0.
\end{align}

In the matrices (\ref{matrix_aniso_0} - \ref{matrix_aniso_3}) we put $u_z = \vartheta = 0$ and denote $A^{i}_{\textrm{aniso}}|_{u_y = \vartheta = 0} = A^{i}_{\perp}$. We consider a solution of equation (\ref{eq_aniso}) in the following form
\begin{equation}
\textbf{W}_\perp(t, x, y, z) = \textbf{Y}_\perp(x) e^{-i(\omega t +ly + mz)}.
\end{equation}
After Laplace transformation for the amplitude vector $\textbf{Y}_\perp$ we have
\begin{equation}
A_\perp \widehat{\textbf{Y}}_\perp (m) = \pm i A_{\perp 1} \textbf{Y}_\perp(0)
\label{eq_perp_1}
\end{equation}
in half-spaces $\Lambda_{\pm}$ respectively, where $A_\perp = \omega A_{\perp 0} + k A_{\perp 1} + l A_{\perp 2} + m A_{\perp 3}$.

Equation (\ref{eq_perp_1}) leads to a characteristic equation $\Det A_\perp = 0$, which reads
\begin{align}
&\Omega \Big[\big((1+c_{s\perp}^2) kv + (F_1 - F_2)\omega\big) \big( \Omega^2 - c_{s\perp}^2(k + v\omega)^2\big) + \nonumber \\ & + (1-2c_{s\perp}^2)(1-v^2) \big((5c_{s\perp}^2 - 3)kv - (F_1 + F_2)w\big) - \nonumber \\ &- c_{s\perp}^2(1-v^2) \big((1+c_{s\perp}^2) kv + (F_1 - F_2)\omega\big)l^2\Big] = 0,
\label{char_perp}
\end{align}
where
\begin{align}
\Omega &= \omega + kv, \\
F_1 &= 2(1 -  c_{s\perp}^2),\\
F_2 &= v^2(1-3 c_{s\perp}^2)
\end{align}
and transverse speed of sound is
\begin{align}
c_{s\perp}^2 = \frac{R_\perp}{3R}.
\end{align}

It can be seen that the only one solution $\Omega = 0$ remains compared to the isotropic and the longitudinal cases. The second one has been transformed into the root of a cubic equation. This feature arises because the anisotropy direction is distinguished, even though the unperturbed problem contains only the $Ox$ axis.

Unfortunately, the cubic equation for $k$ cannot be factorized in a simple way. Using the Cardano formula, one can find the roots of the equation. The roots were studied graphically. For a better interpretation, we move from velocities to the parameter $\sigma = P^{'}/P$. For the velocities $v, v^{'}$ in the regions $\Lambda_{\mp}$, respectively, we have \cite{Kovalenko:2022}:
\begin{align}
v = \sqrt{\frac{R_\perp (3\sigma R + R_\perp)}{3R(\sigma R_\perp + 3R)}}, \ \ \ \ \ v^{'} = \sqrt{\frac{R_\perp (\sigma R_\perp + 3R)}{3R(3\sigma R + R_\perp)}}.
\end{align}
Then we can separately investigate the roots for the half-spaces $\Lambda_{\pm}$. It was found that the imaginary parts of the roots have the same signs as in the longitudinal case. The value of the variables $l, m$ does not affect the final sign of the root in any way, nor does the value of the of the real and imaginary parts of $\omega$. Thus, for the root $k_0 = -v/\omega$ and the roots of the cubic equation $k_{1,2,3}$ one finds
\begin{align}
\Real \omega  \geqslant 0 , \ \Image \omega > 0 &\Leftrightarrow \Real k_{0}  \leqslant 0 , \ \Image k_{0} < 0 \ \ \text{in} \ \Lambda_{\pm},\\
\label{conditions0_perp}
\Real \omega  \geqslant 0 , \ \Image \omega > 0 &\Leftrightarrow \Real k_{1}  \leqslant 0 , \ \Image k_{1} < 0 \ \ \text{in} \ \Lambda_{-},\\
\Real \omega  \geqslant 0 , \ \Image \omega > 0 &\Leftrightarrow \Real k_{1}  \leqslant 0 , \ \Image k_{1} < 0 \ \ \text{in} \ \Lambda_{+},\\
\Real \omega  \geqslant 0 , \ \Image \omega > 0 &\Leftrightarrow \Real k_{2}  \leqslant 0 , \ \Image k_{2} < 0 \ \ \text{in} \ \Lambda_{-},\\
\Real \omega  \geqslant 0 , \ \Image \omega > 0 &\Leftrightarrow \Real k_{2}  \leqslant 0 , \ \Image k_{2} < 0 \ \ \text{in} \ \Lambda_{+},\\
\Real \omega  \geqslant 0 , \ \Image \omega > 0 &\Leftrightarrow \Real k_{3}  \leqslant 0 , \ \Image k_{3} < 0 \ \ \text{in} \ \Lambda_{-},\\
\Real \omega  \geqslant 0 , \ \Image \omega > 0 &\Leftrightarrow \Real k_{3}  \geqslant 0 , \ \Image k_{3} > 0 \ \ \text{in} \ \Lambda_{+}.
\label{conditions_perp}
\end{align}

Since we now have four different roots $k_i$ of the equation (\ref{char_perp}), the following decomposition for the vector $\textbf{Y}_\perp(z)$ is valid:
\begin{equation}
\widehat{\textbf{Y}}_\perp (m) = \frac{\textbf{C}_{0}}{k - k_0} +  \frac{\textbf{C}_{1}}{k - k_1} + \frac{\textbf{C}_{2}}{k - k_2} + \frac{\textbf{C}_{3}}{k - k_3}.
\label{decom_perp}
\end{equation}

Using the inverse Laplace transform and inequalities (\ref{conditions_perp}), one can conclude that in order to satisfy the boundary condition $\delta u_x, \delta u_y, \delta \vartheta, \delta P \rightarrow 0 $ with $x \longrightarrow \pm \infty$ we must put $\textbf{C}_{i} = 0, i = 1,2,3$ in $\Lambda_{-}$ and $\textbf{C}_{3 } = 0$ in $\Lambda_{+}$.

Frome the matching condition $\mu T^{\mu\nu} = N_\mu T^{'\mu\nu}$ he have the following unperturbed equations
\begin{align}
(R_\perp - R_\parallel)(P\sinh(2\vartheta) - P^{'}\sinh(2\vartheta^{'})) &= 0, \\
(P - P^{'})R_\perp - P(R_\perp + R_\parallel)\cosh(2\vartheta) + P^{'}(R_\perp + R_\parallel)\cosh(2\vartheta^{'}) &= 0.
\label{matching_eqs_perp}
\end{align}
from which one obtains
\begin{align}
v v^{'} = \frac{R_\perp}{3R}.
\end{align}

Since we found $\textbf{Y}_\perp(0) = 0$ in half-space $\Lambda_{-}$ then for the velocities we have
\begin{align}
U^\mu &= (u_0, u_x, 0, 0),
\label{perturbed_perp_1} \\
Z^\mu &= (0, 0, 0, 1), \\
U^{'\mu} &= (u_0^{'}, u_x^{'} \delta u_x^{'}, \delta u_y^{'}, \delta\vartheta^{'} u_0^{'}),\\
Z^{'\mu} &= (\delta\vartheta^{'}, 0, 0, 1).
\label{perturbed_perp_4}
\end{align}

Sabstituting the perturbed vectors (\ref{perturbed_perp_1} - \ref{perturbed_perp_4}) to the energy-momentum tensor and the perturbed normal vector $N_\mu = (i \eta \omega, 1, i\eta l, i\eta m)$ into the matching equation one obtains
\begin{align}
Y^0_\perp(0) &= \delta P^{'}  = 2 i \eta \omega P^{'} \frac{R_1 R_2 u_x^{'} u_0^{'} }{3 R R_\perp },
\\
Y^1_\perp(0) &=\delta u_x^{'} = - i \eta k \frac{ R_2 (R_\parallel + R_1 u_x^{'2}) u_0^{'} }{3 R R_\perp },
\\
Y^2_\perp(0) &=\delta u_y^{'} = -i \eta l \frac{ R_\perp - R_2 u_x^{'2} }{ 3 R u_x^{'} },
\\
Y^3_\perp(0) &= \delta \vartheta^{'} = - i \eta m \frac{ R_\parallel (R_\perp - R_2 u_x^{'2}) }{ 3 R_\perp R u_x^{'} u_0^{'}}.
\end{align}

We have all the inputs to solve the equation (\ref{eq_perp_1}) using Cramer's rule. One can obtain formulae for all determinants $\Delta_i$. However, in the transverse case, we will not substitute the roots of the characteristic equation (\ref{char_perp}) into the vector decomposition (\ref{decom_perp}), since the roots are solutions of the cubic equation. Therefore, we will not obtain expressions for constant vectors $\textbf{C}_i$, which do not depend on $k$, and the equation $\textbf{C}_3 = 0$ itself. However, instead of it, it is sufficient for us if all the equations $\Delta_i = 0$, where we consider $k = k_3$, which obeys the conditions (\ref{conditions_perp}).

Denote $M_\perp = A_1^{-1} A_\perp$, then the characteristic equation is equivalent to $\det M_\perp = 0$. It can be verified that that the equation $\Delta_0 = 0$ can be obtained as a linear combination of equations $\Delta_2 = 0$ and $M_\perp$. Moreover it can be shown that the equation $\Delta_3 = 0$ is contained in $\Delta_1^{*}  = 0$ and $\Delta_2 = 0$, where $\Delta_1^{*} $ the combination of $\Delta_1$ and $(c_{s\perp}^2k + \omega v) M_\perp$. Therefore we should consider the equation $\Delta_3 = 0$ which is
\begin{align}
\Omega (w^2 - k^2) = 0,
\end{align}
that in terms of $\varphi$ reads
\begin{align}
\varphi^2 - 2v^{'}\varphi - (1-v^{'2}) = 0.
\label{common_part}
\end{align}
We have obtained the expression which is identical to the equations (\ref{cond}) and (\ref{cond_parallel}).

It should be noted that the equations $\det M_\perp = 0, \Delta_j = 0, j = 1,2,3$ are not reducible to each other only by linear transformations, but it can be shown that their combinations with factors depending on $w, k, m$ transform the equations into each other, thereby highlighting the common part. Thus equations contain not only the common part (\ref{common_part}). In particular, the equation $\Delta_3 = 0$ (and therefore $\Delta_0$ = 0) has the solution
\begin{align}
\varphi = - \frac{v^{'}(3c_{s\perp}^2 - 1)(1-v^{'2})}{2(1-c_{s\perp}^2)+v^{'2}(3c_{s\perp}^2 - 1)},
\end{align}
which is correct for all equation only if $c^2_{s\perp} = 1/2$, i. e. $\xi \rightarrow \infty$.

The last step is to determine the domain of $\varphi$ from the characteristic equation with conditions (\ref{conditions_perp}). In the transverse case, the analysis is carried out differently from in the isotropic case (as well as longitudinal case) and is presented in the Appendix B. It was found that the roots of equation (\ref{common_part}) do not satisfy characteristic equation. Thus, the shock wave in the transverse case is stable.

\subsection{Stability of the shock wave insident at an arbitrary polar angle}

For an arbitrary polar angle $\alpha$ the normal vector takes the form $N^\mu=(0, \sin\alpha, 0 \cos\alpha)$. It is assumed that the upstream flow moves with the velocity $v$, where $v_x = v \sin\alpha, \ v_z = v \cos\alpha$. Behind the shock wave, the downstream flow moves with the velocity $v^{'}$, where $v^{'} _x = v^{'} \sin\alpha^{'}, \ v^{'}_z = v^{'} \cos\alpha^{'}$.

\begin{figure}[h]
\center{\includegraphics[width=0.4\linewidth]{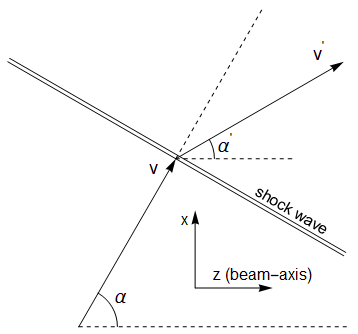}}
\caption{\small Transformation of flow velocity by the shock wave front. Upstream flow moves with velocity \(v\) at an angle \(\alpha\) to the direction of anisotropy (beam-axis) and downstream flow moves with velocity \(v^{'}\) at an angle \(\alpha^{'}\) to the same direction.}
\label{scheme}
\end{figure}

Previously, the properties of such a class of shock waves with constant anisotropy ($\xi = \xi^{'}$) were considered \cite{Kovalenko:2022}. The solutions of the equations $N_\mu T^{\mu\nu} = N_\mu T^{'\mu\nu}$ were obtained by numerical methods due to their analytical unsolvability in the general case. Under certain conditions, for example, for $\alpha = \pi/4$, one can obtain an polynomial of the fifth degree in $v$, and in the case of $\alpha^{'} = 0$, the system of equations is solved analytically. However, we will not consider particular solutions and will carry out the study numerically.

By introducing a harmonic perturbation to the discontinuity surface, we must now take into account the polar angle $\alpha$. The equation of the perturbed surface reads
\begin{equation}
f(t,x,y,z) = x \sin\alpha + z \cos\alpha - e^{-i \big[\omega t + k(x \cos\alpha - z \sin\alpha) + l y\big]} = 0.
\label{dist_surface_alpha}
\end{equation}
It is convenient to move to the coordinate system $\tilde{x}, \tilde{z}$, where the $O\tilde{z}$ axis is directed along the normal $N^\mu$. The matrix defining such a transformation has the form
\begin{equation}
O = 
\begin{pmatrix}
0 & 0 & 0 & 0 \\
0 & \cos\alpha &0 & -\sin\alpha \\
0  & 0 & 0 & 0  \\
0 & \sin\alpha & 0 & \cos\alpha
\end{pmatrix}.
\end{equation}
Coordinate transformations $x, z \rightarrow \tilde{x}, \tilde{z}$ are
\begin{align*}
\tilde{x} &= x \cos\alpha - z \sin\alpha, \\
\tilde{z} &= x \sin\alpha + z \cos\alpha.
\end{align*}
In this coordinate system one finds for the equation (\ref{dist_surface_alpha})
\begin{equation}
f(t,\tilde{x},y,\tilde{z}) = \tilde{z} - e^{-i (\omega t + k\tilde{x} + l y)} = 0.
\label{dist_surface_alpha2}
\end{equation} 

The equations for the energy-momentum tensor in the transformed coordinates are
\begin{equation}
\tilde{\partial}_{\mu}\tilde{T}^{\mu\nu} = 0,
\label{main_eq_general}
\end{equation} 
where $\tilde{T}^{\mu\nu} = O_{\gamma \mu}O_{\lambda \nu}T^{\gamma\lambda}$ and $\tilde{\partial}^{\mu} = O_{\gamma \mu}\partial^{\gamma}$.

The isotropic pressure $P_{\textrm{iso}}$, components of 4-velocity vector $u_x, u_y$ and longitudinal rapidity are linearized according to the formulae (\ref{r1} - \ref{r3}). The linearization of the equations (\ref{main_eq_general}) can be represented as
\begin{equation}
\tilde{A}^\mu \tilde{\partial}_{\mu} \textbf{W} = 0,
\end{equation} 
where the vector $\textbf{W}$ contains the expansion gradients of the quantities defined in (\ref{r1} - \ref{r3}) and have the form
\begin{equation}
\textbf{W} (t,\tilde{x},y,\tilde{z}) = \textbf{Y}(\tilde{z})e^{-i (\omega t + k\tilde{x} + l y)}.
\end{equation}

\begin{figure}[h]
\center{\includegraphics[width=0.7\linewidth]{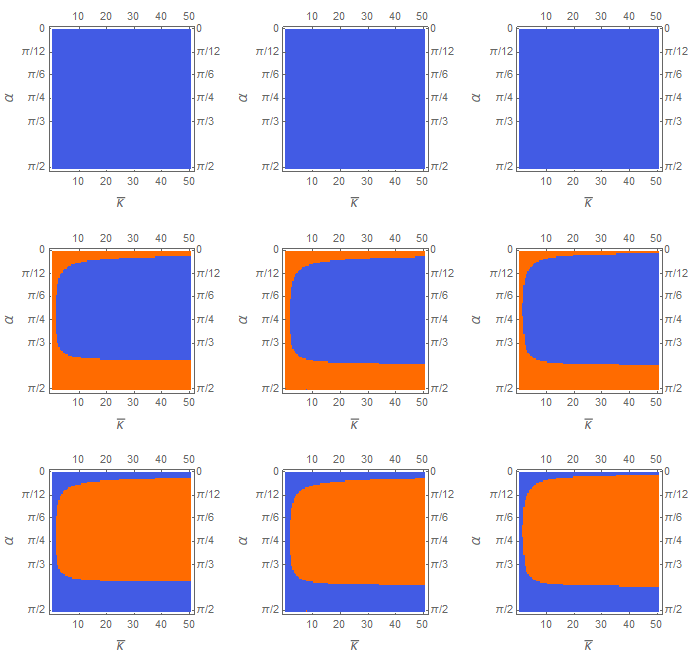}}
\caption{\small Graphs of the imaginary parts of the roots of the characteristic equation in half-space $\Lambda_{+}$ as a function of the polar angle $\alpha$ and the anisotropy ratio $\overline{\kappa}$ for $k = 2, \ l = 2,\ \omega = 2 +2i$. Blue means negative area, orange means positive. The rows correspond to the roots $m_1, m_2, m_3$ and columns correspond to the cases $\sigma = 2,\ \sigma = 10,\ \sigma = 20$ respectively.}
\label{general_im}
\end{figure}

\begin{figure}[h]
\center{\includegraphics[width=0.7\linewidth]{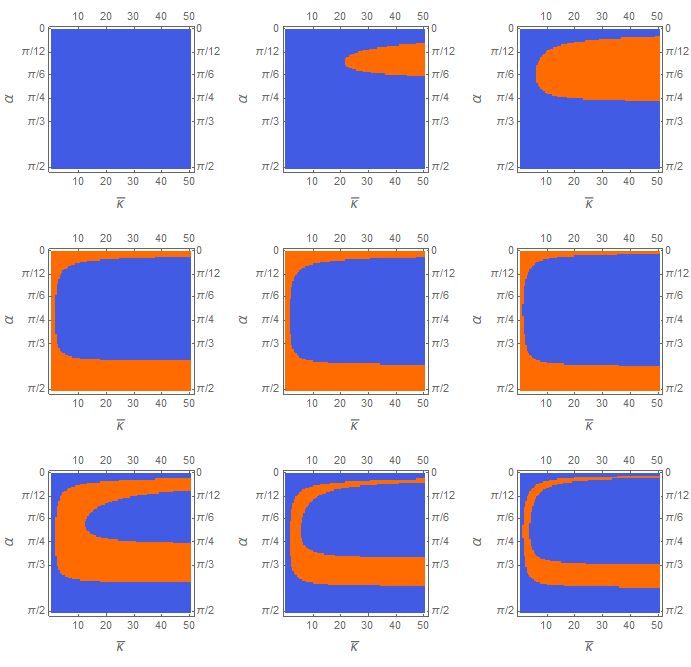}}
\caption{\small Graphs of the real parts of the roots of the characteristic equation in half-space $\Lambda_{+}$ as a function of the polar angle $\alpha$ and the anisotropy ratio $\overline{\kappa}$ for $k = 2, \ l = 2,\ \omega = 2 +2i$. Blue means negative area, orange means positive. The rows correspond to the roots $m_1, m_2, m_3$ and columns correspond to the cases $\sigma = 2,\ \sigma = 10, \ \sigma = 20$ respectively.}
\label{general_re}
\end{figure}

The further sequence of steps is similar to that in the longitudinal case. Applying the Laplace transform to the amplitude vector $\textbf{Y}$ and introducing the variable $m$ gives
\begin{equation}
\tilde{A}\widehat{\textbf{Y}} (m) = \pm i \tilde{A}_{3} \textbf{Y}(0), \ \ \text{in} \ \Lambda_{\pm}.
\label{eq_general_1}
\end{equation}
The corresponding characteristic equation $\det\tilde{A} = 0$ can be solved with respect to $m$. To avoid loss of accuracy and speed of the solution, we pass from the anisotropy parameter $\xi$ to the ratio $\overline{\kappa} = R_\perp(\xi)/R_\parallel(\xi)$. The roots of this characteristic equation are
\begin{equation}
m_0 = - \frac{w - k v^{'} \sin (\alpha - \alpha^{'})}{m v^{'} \cos (\alpha - \alpha^{'})}
\end{equation}
and the three roots of the cubic equation, which we will consider graphically (Figures \ref{general_im} - \ref{general_re}).

For the half-space $\Lambda_{-}$ one can obtain that for all roots $\Image m < 0$, therefore we will not consider this case in detail. As can be seen from the graphs for  $\Lambda_{+}$, the imaginary part of $m$ is negative for the two roots. Moreover, these solutions are deformed in such a way that the existence of a solution with $\Image m > 0$ is possible in the entire phase space. The root $m_1$ shows the same behavior as in longitudinal and transverse cases. The real part of the root $m_2$ repeats the contours of the graphs of the imaginary parts. For this root we again have $\Real m > 0$ if $\Image m > 0, \ \Real \omega > 0$. However, for the root $m_3$ this pattern is violated - here one can see an appearance of a region where $\Real m < 0$ for $\Image m > 0, \ \Real \omega > 0$. We denote this region as $\mathfrak{D}$. The size of this region also depends on $\omega, k, l$, but the region $\mathfrak{D}$ does not completely vanish.

Further, numerically solving the matching equation $N_\mu T^{\mu\nu} = N_\mu T^{'\mu\nu}$ we can find the solution for the vector $\textbf{Y}(0)$. By Cramer's rule one finds a solution to the equation (\ref{eq_general_1}) of the form
\begin{equation}
\widehat{\textbf{Y}} (m) = \Bigg(\frac{\Delta_0}{\Delta_{\tilde{M}}}, \frac{\Delta_1}{\Delta_{\tilde{M}}}, \frac{\Delta_2}{\Delta_{\tilde{M}}}, \frac{\Delta_3}{\Delta_{\tilde{M}}} \Bigg),
\end{equation}
where $\tilde{M} = \tilde{A}_3^{-1} \tilde{A}$. We are interested in the mode for which $\Image \omega > 0$ and $\Image m > 0$. The real parts of $\Real \omega, \ \Real m$, as can be seen from the graphs, can take different signs depending on the different regions of the phase space and the variables $k, l$.

It can be found that the equation $\Delta_0 = 0$ is a linear combination of the characteristic equation $\Delta_{\tilde{M}} = 0$ (or $\Delta_{\tilde{A}} = 0$) and the equation $\Delta_2 = 0$. It can be shown numerically that the solution of the system of equations
\begin{equation}
 \begin{cases}
   \Delta_{\tilde{M}} (\omega, k ,m, l) = 0, \\
   \Delta_2(\omega, k ,m, l) = 0.
 \end{cases}
\label{system_general}
\end{equation}
is also a solution of the equations $\Delta_1 = 0, \Delta_3 = 0$. Therefore, it is possible to confine ourselves to considering only the system (\ref{system_general}).

We introduce the following variables
\begin{align}
\overline{x} = \frac{\omega_R}{m_R}, \ \ \ \ \overline{y} = \frac{\omega_I}{m_I}, \ \ \ \ r = \frac{m_R}{m_I}, \ \ \ \ h = \frac{k}{m_I}.
\end{align}
Since $\Image \omega > 0$ and $\Image m > 0$ we have $\overline{y} > 0$, and for $\overline{x}, r, h$ there exist four cases depending on the signs of $m_R$ and $k$. The system of equations (\ref{system_general}) is divided into four equations (for real and imaginary parts) with unknown $\overline{x}, \overline{y}, r, h, l$. We can write it as
\begin{equation}
 \begin{cases}
   \Real G_M^{(1)}(\overline{x}, \overline{y}, r, h) = \Real G_M^{(2)}(\overline{x}, \overline{y}, r, h) l^2, \\
   \Image G_M^{(1)}(\overline{x}, \overline{y}, r, h) = \Image G_M^{(2)}(\overline{x}, \overline{y}, r, h) l^2, \\
   \Real G_M^{(3)}(\overline{x}, \overline{y}, r, h) l= 0, \\
   \Image G_M^{(3)}(\overline{x}, \overline{y}, r, h) l = 0.
 \end{cases}
\label{system_general2}
\end{equation}
where in the characteristic equation $G_M^{(2)}$ is the coefficient for $l^2$ and $\Delta_2$ can be represented as $G_M^{(3)}(\omega, k ,m) l$. Since $l$ is a real number, we have
\begin{equation}
\Real G_M^{(1)}(\overline{x}, \overline{y}, r, h) \Image G_M^{(2)}(\overline{x}, \overline{y}, r, h) =  \Image G_M^{(1)}(\overline{x}, \overline{y}, r, h)  \Real G_M^{(2)}(\overline{x}, \overline{y}, r, h).
\label{system_ur1}
\end{equation}
This equation is the condition for zeroing the imaginary part of the coefficient for $l^2$ in the characteristic equation. Since $l$ is a real number one has the following inequalities:
\begin{equation}
\frac{\Real G_M^{(1)}(\overline{x}, \overline{y}, r, h)}{\Real G_M^{(2)}(\overline{x}, \overline{y}, r, h)} \geqslant 0 \ \text{    or    } \ \frac{\Image G_M^{(1)}(\overline{x}, \overline{y}, r, h)}{\Image G_M^{(2)}(\overline{x}, \overline{y}, r, h)} \geqslant 0.
\label{system_ur2}
\end{equation}
Thus for $l \neq 0$ we have a system of three equations and inequality
\begin{equation}
 \begin{cases}
\Real G_M^{(1)}(\overline{x}, \overline{y}, r, h) \Image G_M^{(2)}(\overline{x}, \overline{y}, r, h) =  \Image G_M^{(1)}(\overline{x}, \overline{y}, r, h)  \Real G_M^{(2)}(\overline{x}, \overline{y}, r, h), \\
 \Real G_M^{(3)}(\overline{x}, \overline{y}, r, h) = 0, \\
   \Image G_M^{(3)}(\overline{x}, \overline{y}, r, h) = 0,\\
      \frac{\Real G_M^{(1)}(\overline{x}, \overline{y}, r, h)}{\Real G_M^{(2)}(\overline{x}, \overline{y}, r, h)} > 0.
 \end{cases}
\label{system_general3}
\end{equation}
Since $y > 0$, we have to solve three equations for each value of $\overline{y}$ and real $\overline{x}, r, h$, and then see if the inequality is satisfied. If at least one value of $\overline{y}$ satisfies the given system, then the mode of instability exists.

It was found that three equations of the system (\ref{system_general3}) have solutions only for $\overline{x} < 0$. Negative values of $\overline{x}$ is valid because the signs of $\Real \omega$ and $\Real m$ can be different, as discussed above and shown in the Fig. \ref{general_re}.

However, numerical calculations did not lead to any results on the detection of the $\overline{y}$ region where the inequalities (\ref{system_ur2}) hold. Thus it was obtained that for the shock wave incident at an arbitrary polar angle there are no solutions corresponding to the instability mode.

\section{Conclusion}

The linear stability of plane shock waves in ultrarelativistic anisotropic hydrodynamics has been studied. We considered a small harmonic perturbation of the discontinuity surface, which grows exponentially with time. If solutions for perturbed physical quantities vanish at spatial boundaries (at infinity), then an instability mode exists.

The absence of an instability regime was obtained for the solutions of longitudinal and transverse shock waves derived in \cite{Kovalenko:2022}. In the transverse case the influence of the direction of anisotropy was noticed in the solution of the characteristic equation. This effect is expressed in the form of a transformation of a quadratic equation into a cubic one. For both cases, the sign of $\Real \omega$ and $\Image \omega$ uniquely determined the signs of the real and imaginary parts of $m$ (for the longitudinal case) and $k$ (for the transverse case).

The case of a shock wave incident at an arbitrary polar angle $\alpha$ was considered. Two of the three roots of the characteristic equation are mirrored with respect to the sign of $\Image m$. For one of these "mirrored roots", the appearance of a region $\mathfrak{D}$ of the phase space was found, where $\Real m < 0$ with $\Real \omega > 0$ and $\Image \omega > 0$.

The system of equations and inequalities (\ref{system_general3}) was constructed for the input parameters $\xi, \ \sigma, \ \alpha, \ \overline{y}$, the solution of which leads to the existence of an instability mode. However, numerical calculations have shown a violation of inequalities (\ref{system_ur2}), which indicates the absence of an instability mode.

\section{Acknowledgment}

The author are indebted to Professor A. Leonidov for helpful and stimulating discussions in the course of the preparation of this paper.

\appendix
\renewcommand{\theequation}{\thesection.\arabic{equation}}
\section*{Appendix A: Derivation of the domain for $\varphi$ in isotropic case}\label{appenA}

The first step is to write the equation (\ref{char}) in terms of $m$ and $\Omega$:
\begin{equation}
W = \Omega^2 - c_s^2 (m - v^{'2}m + v^{'}\Omega )^2 = c_s^2 (k^2 + l^2) (1-v^{'2}).
\label{char2}
\end{equation}
Define the real and imaginary parts as $\Omega = \Omega_R + i \Omega_I. \ \ m = m_R + i m_I$. Since the right side (\ref{char2}) is real and greater than zero, we have
\begin{align}
\Image W &= v^{'2}(1-c_s^2v^{'2})xy - c_s^2v^{'2}(1-v^{'2})(x + y)- c_s^2(1-v^{'2})^2 = 0, 
\label{imageW}
\\
\Real W &= r^2(x^2v^{'2} - c_s^2(v^{'2}x+1-v^{'2})^2) - (v^{'2}v^2 - c_s^2(v^{'2}y+1-v^{'2})^2) \geqslant 0,
\label{realW}
\end{align}
where
\begin{align}
x = \frac{\Omega_R}{v^{'} m_R}, \ \ \ \ y = \frac{\Omega_I}{v^{'} m_I}, \ \ \ \ r = \frac{m_R}{m_I}.
\end{align}
We are interested in the case of a root $m = m_3$ for which 
\begin{align*}
\Real \omega >0 , \ \Image \omega > 0 &\Leftrightarrow \Real m_{3} > 0 , \ \Image m_{3} > 0 \ \ \text{in} \ \Lambda_{+}.
\end{align*}
Therefore we obtain the following conditions on the introduced variables: $x \geqslant 1, \ y > 1, \ r \in [0, \infty)$.

The expression (\ref{imageW}) represents a hyperbola, which is symmetric in the variables $x, y$. For the function $x(y)$ one finds
\begin{align}
x(y) = \frac{c_s^2v^{'2}(1-v^{'2})y + c_s^2(1-v^{'2})^2}{v^{'2}(1-c_s^2v^{'2})y - c_s^2v^{'2}(1-v^{'2})}.
\label{y_eq}
\end{align}
It can be seen that $x(y)$ decreases as $x$ increases, and for $y > 1$ we obtain
\begin{align*}
x < x(1) = \frac{c_s^2(1-v^{'2})}{v^{'2}(1-c_s^2)}.
\end{align*}

One can define a parametric form $x(s), \ y(s)$, where $x = x(s) = a_0 + a_1 s, \ s \in [s_0, s_*]$ such that $x(s_0) = 1, \ x(s_*) = \frac{ c_s^2(1-v^{'2})}{v^{'2}(1-c_s^2)}$. We require that the denominator (\ref{y_eq}) is proportional to $s$, i.e.
\begin{align}
v^{'2}(1-c_s^2v^{'2})a_0 - c_s^2v^{'2}(1-v^{'2}) = 0,
\end{align}
which gives
\begin{align}
a_0 = \frac{c_s^2(1-v^{'2})}{1-c_s^2v^{'2}}.
\end{align}
The corresponding parameterization for $x(s), \ y(s)$ is
\begin{align}
x &= a_0 +a_1s,
\\
y &= a_0 + \frac{a_2}{s}.
\end{align}
It is natural to require that $a_1 = a_2$, which leads to a system of equations for the boundaries $s_0, \ s_*$ and $a_1$:
\begin{align}
a_0 + a_1 * s_0 &= 1,\\
a_0 + a_1 / s_0 &= \frac{c_s^2(1-v^{'2})}{v^{'2}(1-c_s^2)}, \\
a_0 + a_1 / s_* &= 1,
\end{align}
having the following solution
\begin{align}
a_1 = \frac{c_s(1-v^{'2})}{v^{'}(1-c_s^2v^{'2})}, \ \ s_0 = K, \ \ s_* = 1/K, \ \ K = \frac{v^{'}(1-c_s^2)}{c_s(1-v^{'2})}.
\end{align}
Substitution of the parametric expressions for $x$ and $y$ into the inequality (\ref{realW}) gives
\begin{align}
\Real W &= r^2[a_1^2 (s^2 - 1)] - [a_1^2 (\frac{1}{s^2} - 1)] \geqslant 0.
\label{ineq}
\end{align}
It can be seen that the inequality (\ref{ineq}) is equivalent to $s \geqslant 1$. Thus, for $s$ we have the range $s \in [1, 1/K]$. For $\varphi = \Omega/m$ in terms of $s$ we obtain
\begin{align}
\Real \varphi &= v^{'}a_0 + v^{'}a_1\frac{r^2s + 1/s}{1 + r^2}, \\
\Image \varphi &= v^{'}ra_1\frac{(1/s - s)}{1+ r^2},
\end{align}
with $r \in [0, \infty), \ s \in [1, 1/K]$
These conditions determine the domain of the complex plane $\mathfrak{A}$ for possible values of $\varphi$ that correspond to the instability regime.

For real values of $\varphi$, it is necessary to put $r = 0$, from which one can find the segment of the real line
\begin{align}
v^{'} \leqslant \varphi \leqslant \frac{c_s(1- v^{'2})}{(1-c_sv^{'})}.
\end{align}

\section*{Appendix B: Proof of stability for the transverse anisotropic case}\label{appenB}

Let us define the real and imaginary parts $\omega = \omega_R + i \omega_I. \ \ k = k_R + i k_I$, and introduce the following variables
\begin{align}
x = \frac{\Omega_R}{v^{'} k_R}, \ \ \ \ y = \frac{\Omega_I}{v^{'} k_I}, \ \ \ \ r = \frac{k_R}{k_I}.
\end{align}
We are looking for a specific value of $k$ that satisfies the inequalities
\begin{align*}
\Real \omega >0 , \ \Image \omega > 0 &\Leftrightarrow \Real k_{3} > 0 , \ \Image k_{3} > 0 \ \ \text{in} \ \Lambda_{+},
\end{align*}
therefore $x \geqslant 1, \ y > 1, \ r \in [0, \infty)$.

In the equation (\ref{char_perp}) we denote part that does not depend on $m, l$ as $W$, then we have
\begin{align}
\Real W &= r(-3 L_1 - 2 L_2 y - L_3 y^2 - L_2 x - 2 L_2 xy + 3L_4 xy^2) + (L_1 + L_2x+L_3x^2-L_4x^3)r^3 = \nonumber \\ &= \frac{r}{k_I^2} (1-v^{'2})\Big( c_{s\perp}^2 \big[ (3c_{s\perp}^2 - 1)(1-v^{'2}) + (-2 + v^{'2} + (2-3v^{'2}))x\big]l^2 +  \nonumber \\ &+ (1-2c_{s\perp}^2)\big[ -(3c_{s\perp}^2 - 1)(1-v^{'2}) + (-2 - v^{'2} + (2+3v^{'2}))x\big]m^2\Big) 
\label{realW_perp}
\\
\Image W &= (3 L_1 + 2 L_2 x + L_3 x^2 + L_2 y + 2 L_2 xy - 3L_4 x^2y)r^2 - L_1 - L_2y-L_3y^2+L_4y^3 = \nonumber \\ &= \frac{1}{k_I^2} (1-v^{'2})\Big( c_{s\perp}^2 \big[ (3c_{s\perp}^2 - 1)(1-v^{'2}) + (-2 + v^{'2} + (2-3v^{'2}))y\big]l^2 +  \nonumber \\ &+ (1-2c_{s\perp}^2)\big[ -(3c_{s\perp}^2 - 1)(1-v^{'2}) + (-2 - v^{'2} + (2+3v^{'2}))y\big]m^2\Big),
\label{imageW_perp}
\end{align}
where 
\begin{align}
L_1 &= c_{s\perp}^2 (3c_{s\perp}^2 - 1)(1-v^{'2})^3,\\
L_2 &= c_{s\perp}^2 (1-v^{'2})^2 [2 - 3v^{'2} + c_{s\perp}^2(9v^{'2} - 2)],\\
L_3 &= v^{'2} (1-v^{'2}) [1 +  c_{s\perp}^2(1 - 3v^{'2}) + c_{s\perp}^4(9v^{'2} - 4)],\\
L_4 &= v^{'2} (1-c_{s\perp}^2v^{'2}) [2-v^{'2}-c_{s\perp}^2(2-3v^{'2})].
\end{align}
Since $l, m$ are arbitrary real numbers, it is convenient to include $1/k_I^2$ in the definition of $l, m$.

Consider the imaginary and real parts of $\varphi = (\omega + kv^{'})/k$
\begin{align}
\Real \varphi &=  v^{'}\frac{r^2x + y}{1 + r^2}, \\
\label{real_perp}
\Image \varphi &= v^{'}r\frac{(y -x)}{1+ r^2}.
\end{align}
One should write $r = 0$ or $y = x$, since $\varphi$ takes real values $\varphi_{1,2} = v^{'} \pm 1$. We will consider both cases.

Instead of obtaining the domain of all possible values of $\varphi$ from the characteristic equation, we will substitute the solution $\varphi_{1,2}$ into the characteristic equation. From the roots of $\varphi$ we obtain conditions on $r, y$, which are then applied to the characteristic equation. If in this case the equations (\ref{realW_perp}) and (\ref{imageW_perp}) are valid, then the roots $v^{'} \pm 1$ lie in the proper region.

Consider the case of the solution $r = 0$. From (\ref{real_perp}) one can finds that the roots $v^{'} \pm 1$ lead to condition $y = (v^{'} \pm 1)/v^{'}$. Since $y > 1$, we must choose only $y = (v^{'} + 1)/v^{'}$. Substituting the found solutions into the equations (\ref{realW_perp} - \ref{imageW_perp}) gives 
\begin{align}
\Real W &= 0,
\label{realW_perp2}
\\
\Image W &= - L_1 - L_2\frac{v^{'}+1}{v^{'}}-L_3{(v^{'}+1)^2}{v^{'2}}+L_4{(v^{'}+1)^3}{v^{'3}} = \nonumber \\ &= \frac{1}{v^{'}}(1-v^{'2})\Big( c_{s\perp}^2 \big[ (3c_{s\perp}^2 - 1)(1-v^{'2})v^{'} + (-2 + v^{'2} + (2-3v^{'2}))(v^{'} + 1)\big]l^2 +  \nonumber \\ &+ (1-2c_{s\perp}^2)\big[ -(3c_{s\perp}^2 - 1)(1-v^{'2})v^{'} + (-2 - v^{'2} + (2+3v^{'2}))(v^{'} + 1)\big]m^2\Big).
\label{imageW_perp2}
\end{align}
The last equation provides the following solution for $m^2$
\begin{align}
m^2 = \frac{[(1-c_{s\perp}^2)(1+v^{'}) + c_{s\perp}^2(1-v^{'})l^2][2-v^{'} + c_{s\perp}^2(3v^{'2}-2)]}{(1-2c_{s\perp}^2)(1-v^{'})(-2-v^{'} + c_{s\perp}^2(3v^{'2}+2))}.
\label{eq_for_m}
\end{align}

The first factor in the numerator is greater than zero and $1-2c_{s\perp}^2 \geqslant 0 $ because for the transverse speed of sound we have $1/3 \leqslant c_{s\perp}^2 \leqslant 1/2$. For the second factor in the numerator one can write
\begin{align*}
[2-v^{'} + c_{s\perp}^2(3v^{'2}-2)\Big|_{c^2_{s\perp} = 1/3} &= \frac{4}{3},
\\
[2-v^{'} + c_{s\perp}^2(3v^{'2}-2)\Big|_{c_{s\perp}^2 = 1/2} &=  \frac{v^{'} + 2}{2}.
\end{align*}
Similarly, for the third factor in the denominator we have
\begin{align*}
[-2-v^{'} + c_{s\perp}^2(3v^{'2}+2)\Big|_{c^2_{s\perp} = 1/3} &= -\frac{4}{3},
\\
[-2-v^{'} + c_{s\perp}^2(3v^{'2}+2)\Big|_{c_{s\perp}^2 = 1/2} &=  \frac{v^{'} - 2}{2}.
\end{align*}
One can observe that the entire expression (\ref{eq_for_m}) is less than zero, and since $m$ is a real number, we have a contradiction.

Consider the second case $y = x$, then for $x$ we obtain $x = (v^{'} + 1)/v^{'}$. It can be seen that the equations (\ref{realW_perp} - \ref{imageW_perp}) for $r \neq 0$ can be represented as
\begin{align}
\Real W &= -G_1(x,y) + G_2(x)r^2 = G_3(x, l, m),
\label{realW_perp3}
\\
\Image W &=  G_1(y, x)r^2 - G_2(y) = G_3(y, l, m),
\label{imageW_perp3}
\end{align}
whereas $y = x$ gives
\begin{align}
[-G_1(x,x) + G_2(x) ](r^2 + 1) = 0.
\label{imageW_perp3}
\end{align}
The substitution $x = (v^{'} + 1)/v^{'}$ will not give zero, therefore we have the only solution $r^2 = -1$, which is not in the domain of real $r$.

Thus, we have proved that the instability mode is not observed in the transverse case.

\printbibliography

\end{document}